\documentclass{emulateapj}
\usepackage{apjfonts}
\usepackage{color}

\newcommand{\pivec}{\mbox{\boldmath $\pi$}}
\newcommand{\muvec}{\mbox{\boldmath $\mu$}}
\newcommand{\te}{t_{\rm E}}
\newcommand{\thetae}{\theta_{\rm E}}
\newcommand{\pie}{\pi_{\rm E}}
\newcommand{\pien}{\pi_{{\rm E},N}}
\newcommand{\piee}{\pi_{{\rm E},E}}

\definecolor{darkbrown}{RGB}{139,69,19}

\shorttitle{OGLE-2016-BLG-1112}
\shortauthors{HAN ET AL.}
\begin{document}

\title{OGLE-2014-BLG-1112LB: A Microlensing Brown Dwarf Detected Through 
the Channel of a Gravitational Binary-Lens Event}

\author{
C.~Han$^{1}$, A.~Udalski$^{2,11}$, V.~Bozza$^{3,4,12}$
\\
and\\
M.~K.~Szyma{\'n}ski$^{2}$, I.~Soszy{\'n}ski$^{2}$, J.~Skowron$^{2}$, P.~Mr{\'o}z$^{2}$, 
R.~Poleski$^{2,5}$, P.~Pietrukowicz$^{2}$, S.~Koz{\l}owski$^{2}$, K.~Ulaczyk$^{2}$, 
{\L}. Wyrzykowski$^{2}$\\
(The OGLE Collaboration),\\
S.~Calchi Novati$^{3,6}$, G.~D'Ago$^{7}$, M.~Dominik$^{8}$, 
M.~Hundertmark$^{9}$, U.~G.~Jorgensen$^{10}$, G.~Scarpetta$^{3,7}$\\
(The MiNDSTEp Consortium),\\
}
\affil{$^{1}$  Department of Physics, Chungbuk National University, Cheongju 28644, Republic of Korea}
\affil{$^{2}$  Warsaw University Observatory, Al. Ujazdowskie 4, 00-478 Warszawa, Poland}
\affil{$^{3}$  Dipartimento di Fisica "E. R. Caianiello", Universit\'a di Salerno, Via Giovanni Paolo II, 
               84084 Fisciano (SA), Italy}
\affil{$^{4}$  Istituto Nazionale di Fisica Nucleare, Sezione di Napoli, Via Cintia, 80126 Napoli, Italy}
\affil{$^{5}$  Department of Astronomy, Ohio State University, 140 W. 18th Ave., Columbus, OH 43210, USA}
\affil{$^{6}$  IPAC, Mail Code 100-22, Caltech, 1200 E.~California Blvd., Pasadena, CA 91125}
\affil{$^{7}$  Istituto Internazionale per gli Alti Studi Scientifici (IIASS), Via G.~Pellegrino 19, 
               84019 Vietri sul Mare (SA), Italy}
\affil{$^{8}$  SUPA, School of Physics and Astronomy, University of St.~Andrews, North Haugh, 
               St.~Andrews, KY16  9SS, United Kingdom}
\affil{$^{9}$ Institute and Centre for Star and Planet Formation, University of Copenhagen, 
               {\O}ster Voldgade 5, 1350 Copenhagen, Denmark}
\affil{$^{10}$ Astronomisches Rechen-Institut, Zentrum f\"{u}r Astronomie der Universit\"{a}t, 
               Heidelberg, M\"{o}nchhofstr. 12-14, 69120 Heidelberg, Germany}
\footnotetext[11]{The OGLE Collaboration}
\footnotetext[12]{The MiNDSTEp Consortium}

\begin{abstract}

Due to the nature depending on only the gravitational field, microlensing, in principle, 
provides an important tool to detect faint and even dark brown dwarfs.  However, the 
number of identified brown dwarfs is limited due to the difficulty of the lens mass 
measurement that is needed to check the substellar nature of the lensing object.  In 
this work, we report a microlensing brown dwarf discovered from the analysis of the 
gravitational binary-lens event OGLE-2014-BLG-1112. We identify the brown-dwarf nature 
of the lens companion by measuring the lens mass from the detections of both 
microlens-parallax and finite-source effects. We find that the companion has a mass 
of $(3.03 \pm 0.78)\times 10^{-2}\ M_\odot$ and it is orbiting a solar-type primary 
star with a mass of $1.07 \pm 0.28\ M_\odot$.  The estimated projected separation 
between the lens components is $9.63 \pm 1.33$ au and the distance to the lens is 
$4.84 \pm 0.67$ kpc.  We discuss the usefulness of space-based microlensing observations 
in detecting brown dwarfs through the channel of binary-lens events.
\end{abstract}


\section{Introduction}

Considering that brown dwarfs are formed through a similar process to that of stars 
\citep{Whitworth2007}, it is suspected that the Galaxy would be teeming with brown 
dwarfs. Since microlensing phenomenon occurs due to the gravitational field regardless 
of the brightness of lensing objects, it provides an important tool to detect brown 
dwarfs, especially very faint and dark ones.  Current microlensing surveys detect 
approximately 2000 lensing events each year and an important fraction of them might 
occur due to brown dwarfs.

However, the number of actually identified microlensing brown dwarfs is limited. 
Firm identification of a brown dwarf requires to measure the lens mass in order to 
check that the lens is a substellar object below the hydrogen-burning limit. For 
general lensing events, the only measurable quantity related to the mass of the 
lens is the Einstein time scale $\te$. However, the event time scale results from 
the combination of not only the lens mass but also the relative lens-source transverse 
speed and the distance to the lens. As a result, it is difficult to uniquely determine 
the lens mass based on only $\te$ and this makes it difficult to identify the 
brown-dwarf nature of lens objects for general lensing events. For the unique 
determination of the lens mass, one needs to additionally measure the angular Einstein 
radius $\thetae$ and the microlens parallax $\pie$, which are related to the lens 
mass by the relation \citep{Gould2000}
\begin{equation}
M={\thetae\over \kappa \pie},
\label{eq1}
\end{equation}
where $\kappa=4G/(c^2{\rm au})$. One can measure the angular Einstein radius when a 
lensing event experiences finite-source effects. The microlens parallax can be measured 
by detecting subtle modulations in the lensing light curve produced by the change of the 
observer's position caused by the orbital motion of the Earth around the Sun. For general 
lensing events, unfortunately, the chance to measure both $\thetae$ and $\pie$ is very low.

Although low for general lensing events, the chance to identify a brown-dwarf lens is 
relatively high when events are produced by lenses composed of two masses. There are 
two major reasons for this. First, analysis of a binary-lens event yields an additional 
information of the companion/primary mass ratio $q$ in addition to the time-scale information 
of a single-mass event. Considering that lenses of typical galactic microlensing events are 
low-mass stars, companions of binary lenses with $q<0.1$ are likely to be brown 
dwarf \citep{Shin2012}. Second, the chance to identify the brown-dwarf nature of the lens 
components by measuring both $\thetae$ and $\pie$ is relatively high for a population of 
binary-lens events with long time scales. Most of binary-lens events are identified from 
characteristic features involved with caustics, which represent the positions on the source 
plane at which the lensing magnification of a point source becomes infinite. Due to the high 
magnification gradient around caustics, the light curve of a lensing event during the passage 
over or approach close to the caustics results in deviations caused by finite-source effects, 
enabling one to measure $\thetae$. Furthermore, time scales of binary-lens events tend to be 
longer than those of single-lens events and thus the chance to measure $\pie$ is also higher. 
Due to these reasons, 14 out of the total 16 microlensing brown dwarfs were detected through 
the channel of binary-lens events. See the list of microlensing brown dwarfs presented 
in \citet{Han2016}.

Although useful, detecting brown dwarfs through the binary-lens event channel is still a 
difficult task. Similar to planets, brown dwarfs induce caustic-involved signals in lensing 
light curves. Since planet-induced caustics are usually very small, planetary lensing signals 
in most cases appear as characteristic short-term perturbations to the smooth and symmetric 
form of the light curve produced by the host of the planet \citep{Mao1991, Gould1992a}, making 
it easy to identify the planetary nature of the signal.  
On the other hand, the size of the 
caustic induced by a brown-dwarf companion can be considerable and thus brown-dwarf signals, 
in most cases, cannot be treated as perturbations.\footnote{
We note that perturbations of light curves by binary lenses depend not only on size of the 
caustic but also on the path and size of source. 
Depending on the source trajectory with respect to the caustic and the size of the source star,
the deviation caused by a low-mass binary companion can appear as deviation of 
entire light curves \citep{Ingrosso2009, Ingrosso2011}.  
}
As a result, it is difficult to distinguish 
brown-dwarf binary-lens events from those produced by binaries with roughly equal masses 
\citep{Gaudi2003}.  This implies that finding brown-dwarf events requires detailed analyses 
of numerous binary-lens events which comprise $\sim 10\%$ of all events that are being detected 
with a rate of $\sim 2000\ {\rm yr}^{-1}$. Due to the diversity and complexity of light curves 
that are described by many parameters, modeling light curves of binary-lens events requires a 
complex procedure of analysis. Furthermore, describing caustic-involved features in binary-lens 
light curves requires a numerical approach which demands considerable computing power. As a 
result, the difficulty in analyzing binary-lens events poses an important obstacle in finding 
brown dwarfs via microlensing.

In this paper, we report the microlensing discovery of a binary system that is composed of a 
brown dwarf and a solar type star. The low-mass ratio between the lens components was found 
from the analyses of anomalous events detected in 2014 microlensing season. The angular 
Einstein radius is measured thanks to the good coverage of the caustic-approach region. 
The long time scale of the event combined with extended baseline data  enable us to securely 
measure the microlens parallax. By measuring both $\thetae$ and $\pie$, we uniquely determine 
the lens mass and identify the brown-dwarf nature of the companion.

The paper is organized as follows.  In Section 2, we describe observations of the microlensing 
event analyzed in this work and the data obtained from the observations.  In Section 3, we 
explain the detailed analysis procedure to interpret the observed data and present the best-fit 
model of the lensing event.  In Section 4, we present the physical parameters of the lens 
including the mass and distance.  In Section 5, we discuss the usefulness of space-based lensing 
observations in detecting microlensing brown dwarfs.  We summarize the result and conclude in 
Section 6.

\begin{figure}
\includegraphics[width=\columnwidth]{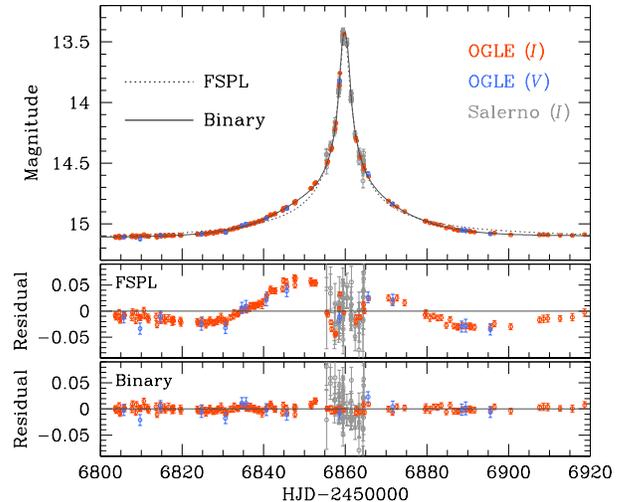}
\caption{
Light curve of the microlensing event OGLE-2014-BLG-1112. 
In the top panel, the solid curve superposed on the data points represents the best-fit 
binary-lens model, while the dotted curve is the finite-source point-lens (FSPL) model.
The middle and bottom panels show the residual from the FSPL and binary-lens models, 
respectively.
}
\label{fig:one}
\end{figure}

\section{Data}

The brown dwarf was discovered from the analysis of the microlensing event OGLE-2014-BLG-1112. 
The event occurred on a star that is located toward the Galactic bulge field. The equatorial 
coordinates of the lensed star (source) are $({\rm RA}, {\rm DEC})_{\rm J2000}=(18^{\rm h} 08^{\rm m} 
36^{\rm s}\hskip-2pt.31, -28^\circ 39' 56.9'')$, which correspond to the Galactic coordinates 
$(l,b)=(2.76^\circ, -4.22^\circ)$. The lensing event was found by the Early Warning 
System of the Optical Gravitational Lensing Experiment \citep[OGLE:][]{Udalski2015a}
that has conducted a microlensing survey since 1992 using the 1.3m telescope at Las Campanas 
Observatory in Chile. The event was also in the footprint of the Microlensing Observations in 
Astrophysics (MOA) survey where the event was dubbed as MOA-2014-BLG-368. We also note that 
there exist data taken by follow-up observations. However, we do not use the MOA and follow-up 
data other than those taken from the Salerno University Observatory because (1) the coverage by 
the OGLE data is dense enough and (2) the photometry quality of the MOA and follow-up data is 
not adequately good enough to measure subtle higher-order effects.  We use the Salerno data 
because they cover the peak region of the light curve.  To be noted is that the lensing-induced 
brightening of the source star started before the 2014 bulge season, lasted throughout the 
season, and continued after the season.

Figure~\ref{fig:one} shows the light curve of the lensing event. At first sight, the light 
curve appears to be that of a point-source point-lens (PSPL) event with a smooth and symmetric 
shape. The light curve of a PSPL event is described by
\begin{equation}
F=AF_s + F_b;\qquad A={u^2+2 \over u\sqrt{u^2+4}},
\label{eq2}
\end{equation}
where $u=\{ [(t-t_0)/\te]^2+u_0^2\}^{1/2}$ 
represents the lens-source separation normalized to the angular Einstein radius $\thetae$, 
$u_0$ is the separation at the moment of the closest lens-source approach $t_0$, $\te$ is the 
Einstein time scale, and $F_s$ and $F_b$ represent the fluxes from the source and blended light, 
respectively.  A close look at the light curve, however, shows a smooth deviation from the PSPL 
model as shown in the residual presented in the middle panel of Figure~\ref{fig:one}.  We note 
that the deviation persisted throughout the event.

In our analysis, we use data acquired from 6-year OGLE observations conducted during 2010 -- 
2015 seasons plus the Salerno data obtained from follow-up observations near the peak of the 
event.  We include OGLE data taken before (2010 to 2013 seasons) and after the lensing 
magnification (2015 season) for the secure measurement of the baseline flux.  The OGLE data 
are composed of two sets taken in $I$ and $V$ bands. The $I$-band data set is composed of 
1557 data points. The $V$-band data, which is composed of 101 data points, are used mainly 
to constrain the source star. Photometry of the OGLE data was conducted using the OGLE pipeline 
\citep{Udalski2003} that is customized based on the Difference Imaging Analysis 
\citep{Alard1998, Wozniak2000}. The Salerno data are reduced by a PSF-fitting pipeline 
developed locally.

Error bars estimated by the pipeline, $\sigma_0$, are 
readjusted following the usual procedure described in \citet{Yee2012}, i.e. 
\begin{equation}
\sigma=k(\sigma_0^2+\sigma_{\rm min}^2)^{1/2},
\label{eq3}
\end{equation}
where the factor $\sigma_{\rm min}$ is used to make the error bars to be consistent with 
the scatter of data and the other factor $k$ is used to make $\chi^2/{\rm dof}=1$. In 
Table~\ref{table:one}, we list the error-bar correction factors $k$ and $\sigma_{\rm min}$ 
along with the number of data points, $N_{\rm data}$.

\begin{deluxetable}{lrcc}
\tablecaption{Errorbar correction factors \label{table:one}}
\tablewidth{0pt}
\tablehead{
\multicolumn{1}{c}{Data set} &
\multicolumn{1}{r}{$N_{\rm data}$} &
\multicolumn{1}{c}{$k$} &
\multicolumn{1}{c}{$\sigma_{\rm min}$} 
}
\startdata
OGLE ($I$)    &  1557      & 1.534   & 0.002      \\
OGLE ($V$)    &  101       & 0.955   & 0.010      \\       
Salerno ($I$) &   60       & 1.23    & 0.010    
\enddata                                              
\end{deluxetable}

\section{Analysis}

Knowing the deviation of the light curve from the PSPL form, we test various 
interpretations of the deviation. For this, we consider effects that are known to 
cause deviations in lensing light curves.

We start with models under the assumption that the lens is a single mass. 
We first check finite-source effects (FSPL model) which occur when the source 
passes over or approaches very close to the lens.  From FSPL modeling, it is 
found that the deviation cannot be attributed to the finite-source effect 
because the deviation lasted $\gtrsim 120$ days, while the effect of the finite 
source is confined to a short time range. 
We, therefore, check two known causes of long-term deviations.  The first one 
is the microlens parallax effect.  Light curve deviation by the parallax effect 
is caused by the deviation of the relative lens-source motion from rectilinear 
due to the orbital motion of the Earth around the sun (Gould 1992). In addition 
to the single-lensing parameters, i.e.\ $t_0$, $u_0$, and $\te$, consideration 
of the parallax effect requires to include two additional parameters $\pien$ and 
$\piee$, which denote the north and east components of the microlens parallax 
vector $\pivec_{\rm E}$ projected on the sky in the north and east equatorial 
coordinates, respectively.  We also check the possibility that the source is 
composed two stars: `binary-source point-lens' (BSPL) model \citep{Griest1992}. 
Considering the source binarity requires to include 3 additional parameters $t_{0,2}$, 
$u_{0,2}$, and $q_f=F_{s2}/F_{s1}$, which represent the time and impact parameter of 
the closest lens approach to the source companion, and the flux ratio between the 
source stars, respectively \citep{Han1998}. We find that neither of these interpretations 
provides a model that can explain the deviation.

\begin{figure}
\includegraphics[width=\columnwidth]{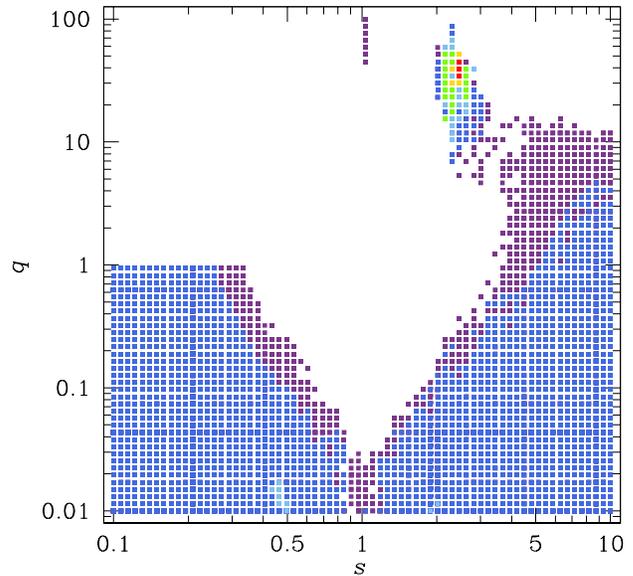}
\caption{
$\Delta\chi^2$ distribution in the $s$-$q$ parameter space obtained from the preliminary 
grid search for lensing solutions. Color coding denotes points in the MCMC chain with $1n\sigma$ 
(red), $2n\sigma$ (yellow), $3n\sigma$ (green), $4n\sigma$ (cyan), $5n\sigma$ (blue), 
and $6n\sigma$ (purple), where $n=15$.
}
\label{fig:two}
\end{figure}

We then check the possibility of the lens binarity. Under the assumption of a rectilinear 
relative lens-source motion, the lensing light curve of a binary-lens event is described 
by 7 principal parameters. Three of these parameters are same as those of a single-lens event, 
i.e.\ $t_0$, $u_0$, and $\te$. Two other parameters are needed to describe the binary lens, 
including the projected separation, $s$, and the mass ratio, $q$, between the lens components. 
The separation is normalized to $\thetae$. The source trajectory angle, $\alpha$, which is 
defined as the angle between the source trajectory and the binary lens axis, is needed to specify 
the source trajectory with respect to the lens. Finally, the normalized source radius $\rho$, 
which is defined as the ratio of the angular source radius $\theta_*$ to the angular Einstein 
radius, i.e.\ $\rho=\theta_*/\thetae$, is needed to describe the deviation in the caustic-crossing 
part of the lensing light curve caused by finite-source effects. For a single lens, the reference 
position on the lens plane is that of the lens itself, but for a binary lens, for which there 
exist two lens components, one needs to define a reference position. We use the barycenter of 
the binary lens as a reference position. Although not apparent, the light curve may be associated 
with caustics, during which the light curve was affected by finite-source effects.  We compute 
finite-source magnifications using the numerical ray-shooting method. 
In this process, we consider the limb-darkening variation of the source star surface with 
the model of the surface brightness profile $S\propto 1-\Gamma(1-3\cos\phi/2)$, where $\Gamma$ 
is the limb-darkening coefficient and $\phi$ is the angle between the line of sight toward the 
source center and the normal to the source surface.  The adopted limb-darkening coefficients, 
$\Gamma_I=0.53$ and $\Gamma_V=0.73$ for the $I$ and $V$ band data sets, respectively, are 
chosen from \citet{Claret2000} considering the source type that is determined based on the 
de-reddened color $(V-I)_0$ and brightness $I_0$. See Section 4 for the detailed procedure 
of $(V-I)_0$  and $I_0$ determinations.

Due to the multiplicity of the lensing parameters and the resulting diversity of light curves, it 
is known that light curves of binary-lensing events are subject to various types of degeneracy 
where different combinations of the lensing parameters result in similar lensing light curves. 
In the preliminary search, therefore, we conduct a grid search for lensing solutions not only to 
find the lensing solution but also to check the existence of degenerate solutions. 
The grid search is conducted in the space of ($s$, $q$, $\alpha$) parameters for which lensing 
magnifications vary sensitively to the small changes of these parameters. The ranges of $s$ and 
$q$ parameters are $-1 < \log s < 1$ and $-4 < \log q < 2$, respectively. The magnification 
variation to the changes of the other lensing parameters is smooth and thus we search for these 
parameters by minimizing $\chi^2$ using a downhill approach. For the downhill approach, we use 
the Markov Chain Monte Carlo (MCMC) method.

\begin{deluxetable}{clcc}
\tablecaption{Comparison of models \label{table:two}}
\tablewidth{0pt}
\tablehead{
\multicolumn{2}{c}{Model} &
\multicolumn{1}{c}{$\chi^2$} 
}
\startdata
PSPL    & (Parallax)                 &  4371.2  \\
FSPL    & (Parallax)                 &  4283.5  \\
BSPL    & (Parallax)                 &  3729.8  \\     
Binary  & (Standard)                 &  1771.8  \\ 
  -     & (Parallax, $u_0<0$)        &  1565.5  \\ 
  -     & (Parallax, $u_0>0$)        &  1563.3  \\
  -     & (Orbit)                    &  1535.4  \\
  -     & (Orbit+parallax, $u_0<0$)  &  1532.5  \\
  -     & (Orbit+parallax, $u_0>0$)  &  1533.1
\enddata                                              
\end{deluxetable}

From the preliminary grid search, we find a unique solution of lensing parameters describing 
most part of the observed light curve. Figure 2 shows the location of the solution on the 
$\Delta\chi^2$ distribution in the $s$-$q$ parameter space obtained from the preliminary grid 
search.  The estimated values of the binary separation and mass ratio are $s\sim 2.4$ and 
$q\sim 40$, respectively.  The value of $s$ indicates that the projected separation between 
the lens components is $\sim 2.4$ times greater than the angular Einstein radius. The fact 
that the mass ratio is greater than unity indicates that the source approached closer to the 
lower-mass lens component.  The large mass ratio of $q\sim 40$ indicates that the companion 
is a low-mass object.

Since the event lasted throughout the whole bulge season, the light curve of the event might 
be affected by long-term higher-order effects, such as the effects of the Earth's orbital 
motion and the orbital motion of the lens itself \citep{Albrow2000, Shin2012, Park2013}. We, 
therefore, conduct additional modeling considering these higher-order effects. In the `parallax' 
and `orbit' models, we separately consider the parallax effect and the lens-orbital motion, 
respectively. In the `orbit+parallax' model, we simultaneously consider both higher-order effects. 
The parameters describing the parallax effect, $\pien$ and $\piee$, are same as those defined for 
single-lens events. To the first order approximation, the lens-orbital effect is described by two 
parameters of $ds/dt$ and $d\alpha/dt$, which represent the change rates of the binary separation 
and the source trajectory angle, respectively. When the parallax effect is considered, we check 
the degeneracy in the parallax parameters between the pair of models with $u_0 < 0$ and $u_0 > 0$. 
This so-called `ecliptic degeneracy' is caused by the mirror symmetry of the source trajectory 
with respect to the binary axis \citep{Skowron2011}. The parameters of the two degenerate 
solutions are in the relation $(u_0, \alpha, \pien) \leftrightarrow -(u_0, \alpha, \pien)$.

\begin{figure}
\includegraphics[width=\columnwidth]{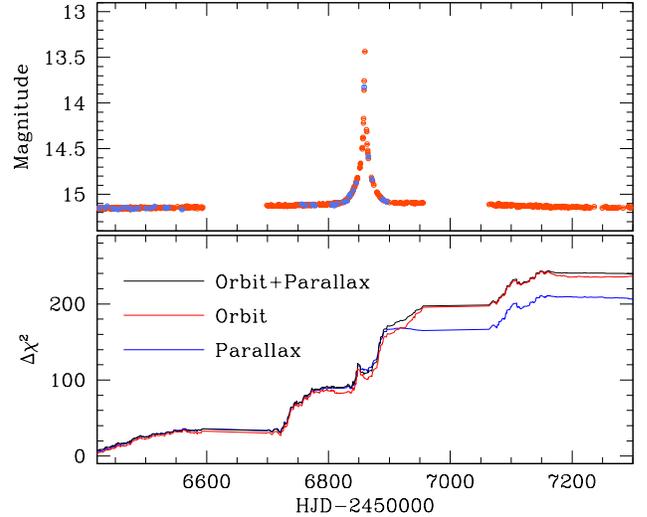}
\caption{
Cumulative $\Delta\chi^2$ distributions of the models with higher-order effects with respect 
to the model without considering the higher-order effects.  The event light curve in the 
upper panel is presented to show the region of $\chi^2$ improvement.
}
\label{fig:three}
\end{figure}

\begin{figure}
\includegraphics[width=\columnwidth]{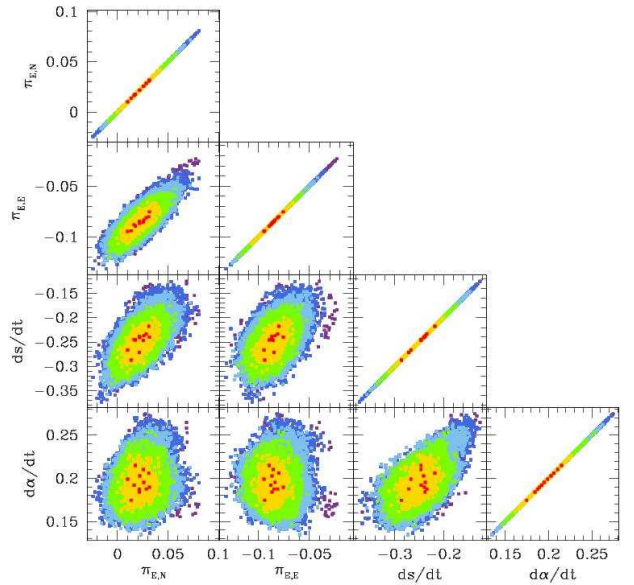}
\caption{
$\Delta\chi^2$ Distributions of higher-order lensing parameters: $\pien$, $\piee$, $ds/dt$, and 
$d\alpha/dt$. Color coding denotes points in the MCMC chain with $1\sigma$ (red), $2\sigma$ (yellow), 
$3\sigma$ (green), $4\sigma$ (cyan), and $5\sigma$ (blue) of the best-fit value. We note that the 
distributions are for ¿orbit+parallax¿ model with $u_0<0$.
}
\label{fig:four}
\end{figure}

In Table~\ref{table:two}, we compare the goodness of fits of the tested models. For comparison, 
we also present the $\chi^2$ values of the PSPL, FSPL, and BSPL models. From the comparison, we 
find that $\chi^2$ difference between the single-mass (PSPL, FSPL, and BSPL) and the binary-lens models 
is $\Delta\chi^2 > 2000$, indicating that the event was produced by a binary object. From the 
comparison of $\chi^2$ values of the individual binary-lens models, we find that the separate 
considerations of the parallax and lens-orbital effects improve the fit by $\sim 208$ and 236, 
respectively, with respect to the model based on the principal binary-lensing parameters 
(`standard model'). We also find that simultaneous consideration of both higher-order effects 
improves the fit by $\sim 239$. Although the $\chi^2$ difference between the `orbit' and 
`orbit+parallax' models is minor, i.e. $\sim 3$, it is found that the microlens-parallax parameters 
with and without considering the lens-orbital effects are slightly different and thus we judge 
that both higher-order effects are important for the precise description of the observed lensing 
light curve. In Figure~\ref{fig:three}, we present the cumulative distribution of $\Delta\chi^2$ 
of the models considering higher-order effects with respect to the model without considering the 
higher-order effects.  As expected from the long-term microlens-parallax and lens-orbital effects, 
the improvement of the fit occurs throughout the event.  In Figure~\ref{fig:four}, we present the 
distributions of $\Delta\chi^2$ in the parameter space of the higher-order lensing parameters. We 
note that although the measured microlens parallax value $\pi_{\rm N}=(\pien^2+\piee^2)^{1/2}=0.092$ 
is small, it is measured with a significant confidence level.

\begin{deluxetable}{lrr}
\tablecaption{Best-fit lensing parameters \label{table:three}}
\tablewidth{0pt}
\tablehead{
\multicolumn{1}{c}{Parameters} &
\multicolumn{1}{c}{$u_0>0$}    &
\multicolumn{1}{c}{$u_0<0$}  
}
\startdata
$t_0$ (HJD')                     &  6892.200 $\pm$ 0.994 &  6890.985 $\pm$ 0.174  \\  
$u_0$                            &  -1.848   $\pm$ 0.011 &  1.861    $\pm$ 0.001  \\    
$\te$ (days)                     &  107.44   $\pm$ 1.10  &  106.37   $\pm$ 0.85   \\     
$s$                              &  2.43     $\pm$ 0.01  &  2.42     $\pm$ 0.01   \\  
$q$                              &  35.37    $\pm$ 1.56  &  47.38    $\pm$ 3.35   \\    
$\alpha$ (rad)                   &  1.404    $\pm$ 0.008 &  -1.411   $\pm$ 0.001  \\ 
$\rho$ ($10^{-3}$)               &  1.14     $\pm$ 0.03  &  1.01     $\pm$ 0.04   \\  
$\pi_{{E},N}$                    &  0.02     $\pm$ 0.01  &  -0.01    $\pm$ 0.01   \\   
$\pi_{{E},E}$                    &  -0.09    $\pm$ 0.01  &  -0.08    $\pm$ 0.01   \\     
$ds/dt$ (${\rm yr}^{-1}$)        &  -0.25    $\pm$ 0.03  &  -0.29    $\pm$ 0.03   \\     
$d\alpha/dt$ (${\rm yr}^{-1}$)   &  0.22     $\pm$ 0.02  &  -0.40    $\pm$ 0.05   \\    
$(F_s/F_b)_{{\rm OGLE},I}$       &  13.49/0.26           &   13.69/ 0.06             
\enddata                                              
\tablecomments{${\rm HJD}'={\rm HJD}-2450000$.}
\end{deluxetable}

\begin{figure}
\includegraphics[width=\columnwidth]{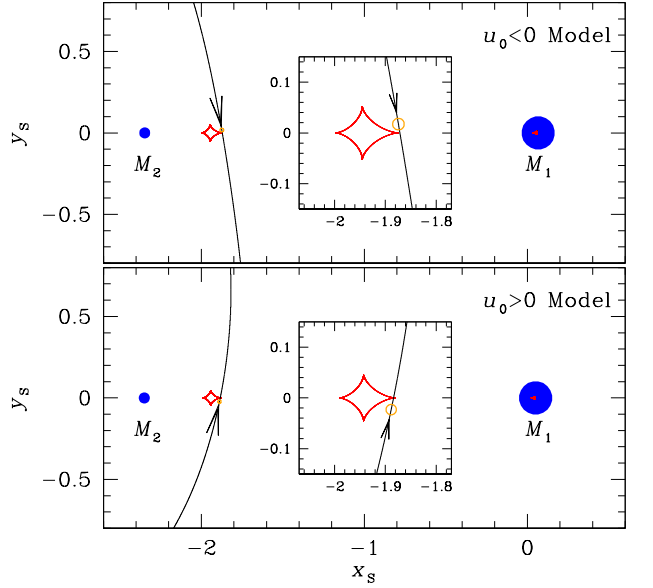}
\caption{
Geometry of the lens system. The curve with an arrow is the source trajectory and the blue dots 
represent the locations of the binary lens components, where $M_1$ and $M_2$ denote the heavier 
and lighter mass components. The cuspy close curve represent the caustic. The inset shows the 
enlarged view of the caustic that is located closer to the lower-mass lens component. All length 
are normalized to the angular Einstein ring radius corresponding to the total mass of the binary 
lens and the coordinates are centered at the barycenter of the binary lens.
}
\label{fig:five}
\end{figure}

In Table~\ref{table:three}, we present the lensing parameters of the best-fit model, i.e.\ 
`orbit+parallax' binary-lens model.  Since the ecliptic degeneracy is very severe (with 
$\Delta\chi^2=0.6$), we present both the $u_0<0$ and $u_0>0$ solutions. In Figure~\ref{fig:five}, 
we also present the lens-system geometry which shows the source trajectory (curve with an arrow) 
with respect to the lens components (marked by blue dots) and the caustic (red cuspy closed curves). 
We note that the upper and lower panels correspond to the $u_0<0$ and $u_0>0$ solutions, respectively. 
It is found that the lens is a wide binary for which two sets of caustics form near the individual 
lens components.  The event was produced by the source trajectory which passed the tip of the caustic 
located close to the lower-mass lens component, $M_2$. Since the source trajectory almost vertically 
passed the binary axis ($\alpha\sim 80^\circ$), the light curve appears to be symmetric with respect 
to the peak that corresponds to the time of the caustic crossing.  In Figure~\ref{fig:one}, we present 
the best-fit model light curve, i.e.\ `orbit+parallax' model with $u_0<0$, superposed on the data 
points. In the bottom panel, we also present the residual of the model. One finds that the residual 
from the FSPL model has gone with the introduction of the lens binarity.

\begin{figure}
\includegraphics[width=\columnwidth]{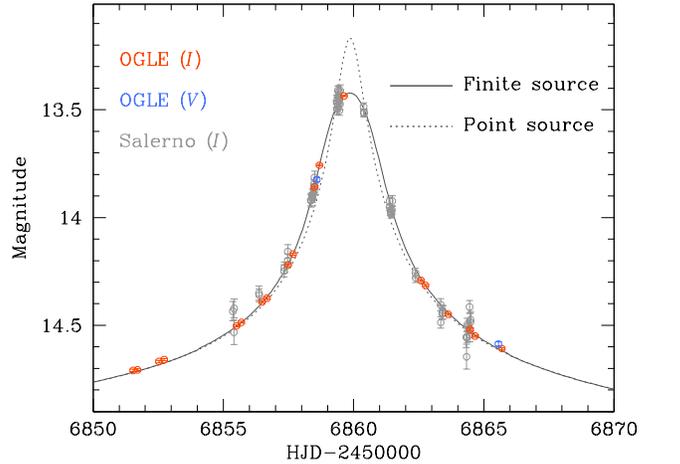}
\caption{
Peak region of the light curve. Dotted and solid curves represent the model light curves with 
a solid and a finite source, respectively.
}
\label{fig:six}
\end{figure}

We note that finite-source effects are clearly detected, although there exists no prominent spike 
caustic-crossing features that commonly appear in binary-lens light curves. This can be seen in 
the peak region of the light curve presented in Figure~\ref{fig:six}, where we plot two model 
light curves with a point (dotted curve) and a finite source (solid curve). The measured values 
of the normalized source radius are $\rho=(1.14 \pm 0.03)\times 10^{-2}$ and 
$(1.01 \pm 0.03)\times 10^{-2}$ for the $u_0<0$ and $u_0>0$ solutions, respectively. In the inset 
of Figure~\ref{fig:five}, we mark the source near the caustic as a circle where the size of the 
circle is scaled to the caustic size. It shows that the source crossed the tip of the caustic. 
We note that finite-source effects could also have been detected 
through specific features of polarization curves if a polarization observation
of the event were conducted \citep{Ingrosso2012, Ingrosso2014}.

\begin{figure}
\includegraphics[width=\columnwidth]{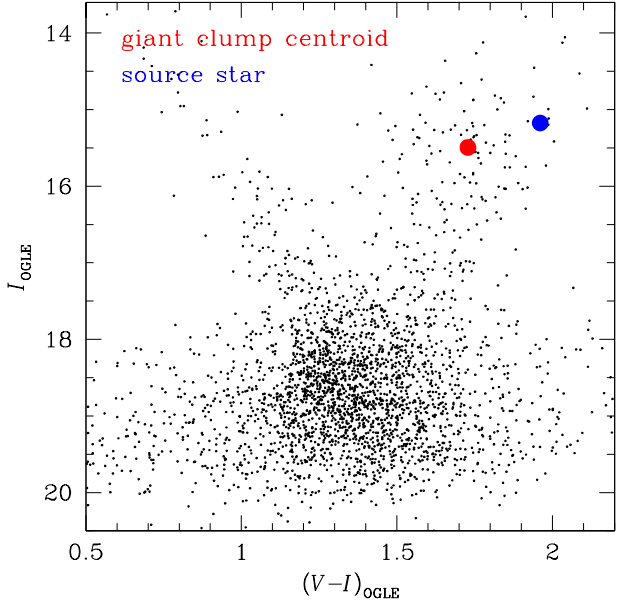}
\caption{
Source location with respect to that of the centroid of giant clump in the color-magnitude diagram.
}
\label{fig:seven}
\end{figure}

\section{Physical Parameters}

While the subtle modulation of the lensing light curve allows us to measure the microlens parallax $\pie$,
detections of the finite-source effect enables us to measure the angular Einstein radius 
$\thetae$, which is the other ingredient for the lens mass determination. The angular Einstein 
radius is related to the normalized source radius $\rho$ and the angular source radius $\theta_*$ by
$\thetae=\theta_*/\rho$, and thus one needs to estimate $\theta_*$ for the $\thetae$ measurement. 
We estimate angular source radius from the de-reddened color and brightness following the 
method of \citet{Yoo2004} using the centroid of bulge giant clump (GC) as a standard candle. 
In this method, the de-reddened color and magnitude of the source stars are determined by
\begin{equation}
(V-I,I)_0=(V-I,I)_{\rm GC} + \Delta (V-I,I),
\label{eq4}
\end{equation}
where $(V-I,I)_{\rm GC}=(1.06, 14.29)$ are the color and magnitude of the GC centroid 
\citep{Bensby2011, Nataf2013} and $\Delta (V-I,I)$ are offsets of the color and magnitude 
of the source star from those of the GC centroid measured in the instrumental (uncalibrated) 
color-magnitude diagram. In Figure~\ref{fig:seven}, we mark the locations of the source and 
centroid of GC in the instrumental color-magnitude diagram of stars in the neighboring region 
around the source star. We find that the dereddened color and magnitude of the source star are 
$(V-I,I)_0=(1.29,13.98)$, indicating that the source is a K-type giant. We then estimate the 
angular source radius by first converting $V-I$ into $V-K$ using the color-color relation 
\citep{Bessell1988} and then estimate $\theta_*$ using the relation between $V-K$ and the 
surface brightness \citep{Kervella2004}.

\begin{deluxetable}{lcc}
\tablecaption{Source parameters \label{table:four}}
\tablewidth{0pt}
\tablehead{
\multicolumn{1}{c}{Quantity} &
\multicolumn{1}{c}{$u_0<0$} &
\multicolumn{1}{c}{$u_0>0$} 
}
\startdata
$\theta_*$ (uas)                         &  9.30 $\pm$ 1.35   &  9.37 $\pm$ 1.36   \\
$\thetae$ (mas)                          &  0.82 $\pm$ 0.12   &  0.93 $\pm$ 0.14   \\
$\mu_{\rm geo}$  (mas ${\rm yr}^{-1}$)   &  2.78 $\pm$ 0.41   &  3.17 $\pm$ 0.48   \\
$\mu_{\rm helio}$ (mas ${\rm yr}^{-1}$)  &  2.52 $\pm$ 0.38   &  2.93 $\pm$ 0.44          
\enddata                                              
\end{deluxetable}

In Table~\ref{table:four}, we present the estimated angular source radius and 
Einstein radius. Also presented are the relative lens-source proper motion measured in the 
geocentric and heliocentric frames. The geocentric proper motion is estimated by 
\begin{equation}
\mu_{\rm geo}={\thetae \over \te}. 
\label{eq5}
\end{equation}
The heliocentric proper motion is measured by
\begin{equation}
\muvec_{\rm helio}=\muvec_{\rm geo} + v_{\oplus,\perp}{\pi_{\rm rel}\over {\rm au}},
\label{eq6}
\end{equation}
where $\muvec_{\rm geo}=[\mu_{\rm geo}(\pien/\pi_{\rm E}),\mu_{\rm geo}(\piee/\pi_{\rm E})]$,  
$\pi_{\rm rel}$ is the relative lens-source parallax, and 
$ v_{\oplus,\perp}$
is the velocity of the Earth projected on the sky at $t_0$.
We note that direction of $\muvec_{\rm helio}$ is same as that of the microlens parallax 
vector $\pivec_{\rm E}$.

\begin{deluxetable}{lcc}
\tablecaption{Physical lens parameters \label{table:five}}
\tablewidth{0pt}
\tablehead{
\multicolumn{1}{c}{Parameter} &
\multicolumn{2}{c}{Model} \\
\multicolumn{1}{c}{} &
\multicolumn{1}{c}{$u_0>0$} &
\multicolumn{1}{c}{$u_0<0$} 
}
\startdata
$M_1$ ($M_\odot$)            & 1.07 $\pm$  0.28     &  1.40  $\pm$ 0.40     \\
$M_2$ ($10^{-2} M_\odot$)    & 3.03 $\pm$  0.78     &  2.95  $\pm$ 0.84     \\
$D_{\rm L}$ (kpc)            & 4.84 $\pm$  0.67     &  4.87  $\pm$ 0.71     \\
$a_\perp$ (au)               & 9.63 $\pm$  1.33     &  10.90 $\pm$ 1.58    \\
$({\rm KE}/{\rm PE})_\perp$  & 0.58                 &  2.00           
\enddata                                              
\end{deluxetable}

By measuring both the microlens parallax and the angular Einstein radius, we can determine 
the mass of the lens by using the relation in Equation (1). One can also determine the 
distance to the lens by using the relation \citep{Gould2000}
\begin{equation}
D_{\rm L}={{\rm au}\over \pie\thetae + \pi_{\rm S}},
\label{eq7}
\end{equation}
where $\pi_{\rm S}={\rm au}/D_{\rm S}$ is the parallax of the source star and $D_{\rm S}$
is the distance to the source star. 
In Table~\ref{table:five}, we present the  determined physical parameters.
The masses of the lens components are estimated by
$M_1=M/(1+q)$ and $M_2=Mq/(1+q)$, where $M=M_1+M_2$.
The projected separation between the lens components is estimated by $a_\perp=s D_{\rm L}\thetae$.
Also presented is the projected kinetic to potential energy ratio which is computed by 
\begin{equation}
\left( {{\rm KE} \over {\rm PE}} \right)_{\perp}=
{(r_{\perp}/{\rm au})^3 \over 8 \pi^{2}(M/M_{\odot})}
\left[ \left( {1 \over s} {ds \over dt}\right)^2 + 
\left({d \alpha \over dt} \right)^2 \right].
\label{eq8}
\end{equation}
In order for the binary lens to be a bound system, the ratio should satisfy the condition 
$({\rm KE}/{\rm PE})_\perp \leq {\rm KE}/{\rm PE} \leq 1$, where {\rm KE}/{\rm PE} is the 
intrinsic energy ratio.  It is found that the energy ratio for the $u_0>0$ model is greater 
than unity, indicating that the solution results in unphysical parameters.  This leaves the 
$u_0<0$ model the only viable solution.

According to the solution, the lens is composed of a solar-type primary star
with a mass of $M_1=1.07 \pm 0.28\ M_\odot$ and a brown-dwarf companion with
a mass of $M_2=(3.03 \pm 0.78)\times 10^{-2}\ M_\odot$.
The companion has a mass less than the hydrogen-burning limit of $\sim 0.08\ M_\odot$ 
\citep{Hayashi1963, Nakano2014}.
The projected separation between the lens components is $a_\perp=9.63 \pm 1.33$ au.
The lens is located at a distance $D_{\rm L}=4.84 \pm 0.67$ kpc.

\section{Discussion}

Before 2014, nearly all microlens mass measurements were based on the `annual parallax', 
which is determined from the modulation of a lensing light curve caused by the annual 
orbital motion of the Earth. A microlens parallax can also be measured by conducting 
simultaneous observations of lensing events from a ground-based observatory and from a 
satellite in solar orbit \citep{Refsdal1966, Gould1994}. Since 2014, a space-based 
microlensing campaign using the {\it Spitzer} telescope, which is in solar orbit with a 
projected separation toward the bulge of $d_{\rm proj}\sim 1$ au, has been conducted 
\citep{Calchi2015}. By successfully measuring microlens parallaxes of various types of 
lensing events,  e.g.\  \citet{Udalski2015b}, \citet{Street2016}, \citet{Yee2015}, 
\citet{Shvartzvald2016}, \citet{Zhu2015}, \citet{Bozza2016}, \citet{Han2016}, the campaign 
demonstrated that space-based parallaxes can be routinely measured for general lensing events. 
In 2016 season, a new microlensing survey has been conducted using the {\it Kepler} satellite, 
which has projected separations spanning $0.07\ {\rm au} < d_{\rm proj} < 0.81\ {\rm au}$ 
over the course of the 2016 season \citep{Henderson2016}. The data collected from the 
campaign, dubbed {\it K2}'s Campaign 9 ({\it K2C9}), are being processed and microlens 
parallaxes of numerous events are expected to be measured. Furthermore, the {\it WFIRST} 
space telescope, which is set to launch in the mid-2020s, will be able to measure space-based 
microlens parallaxes for increased number of lensing events.

Binary events will be important targets for the detections of brown dwarfs in the current 
and future space-based lensing observations. For binary-lens events, angular Einstein 
radii are routinely measurable and thus additional measurement of space-based microlens 
parallaxes will enable to measure lens masses. 
Furthermore, while the microlens-parallax measurement for a single-mass event suffers 
from a well-known two-fold degeneracy \citep{Gould1994}, the microlens parallax of a binary-lens
event is, in general, uniquely determined \citep{Han2017}. 
Being able to routinely measure masses of 
binary lenses, therefore, the number of microlensing brown dwarfs detected through the 
channel of binary-lens events is expected to be greatly increased.

\section{Conclusion}

We reported a microlensing brown dwarf discovered from the analysis of the gravitational 
binary-lens event OGLE-2014-BLG-1112.  We identified the brown-dwarf nature of the lens 
companion by measuring the lens mass from the detections of both microlens-parallax and 
finite-source effects. We found that the companion has a mass of $(3.03 \pm 0.78)\times 
10^{-3}\ M_\odot$ and it is orbiting a solar-type primary star with a mass of 
$1.07 \pm 0.28\ M_\odot$.  The estimated projected separation between the lens components 
was $9.63 \pm 1.33$ au and the distance to the lens was $4.84 \pm 0.67$ kpc.  We discussed 
the usefulness of space-based microlensing observations in detecting brown dwarfs through 
the channel of binary-lens events.

\begin{acknowledgments}
Work by C.~Han was supported by the Creative Research Initiative Program (2009-0081561) of 
National Research Foundation of Korea.  
The OGLE project has received funding from the National Science Centre, Poland, grant 
MAESTRO 2014/14/A/ST9/00121 to A.~Udalski.  OGLE Team thanks Profs.\ M.~Kubiak and 
G.~Pietrzy{\'n}ski, former members of the OGLE team, for their contribution to 
the collection of the OGLE photometric data over the past years.
We acknowledge the high-speed internet service (KREONET)
provided by Korea Institute of Science and Technology Information (KISTI).

\end{acknowledgments}

\end{document}